\documentclass[%
reprint,
showpacs,preprintnumbers,
 amsmath,amssymb,
 aps,
]{revtex4-1}
\usepackage{graphicx}
\usepackage{dcolumn}% Align table columns on decimal point
\usepackage{bm}% bold math
\usepackage{epstopdf}
\usepackage{amsmath}
\usepackage{bm}
\usepackage{mathtools}

\newcommand{\cc}{\mathrm{c\bar{c}}}
\newcommand{\Jpsi}{\mathrm{J/\psi}}
\newcommand{\bb}{\mathrm{b\bar{b}}}
\newcommand{\Y}{\mathrm{\Upsilon}}
\newcommand{\YnS}[1]{{\Y({#1}\mathrm{S})}}
\newcommand{\chib}{\mathrm{\chi_b}}
\newcommand{\chibnP}[1]{{\chib({#1}\mathrm{P})}}

\newcommand{\RAA}[1][]{{R_{AA}^{#1}}}
\newcommand{\RAAQGP}{{\RAA[\text{QGP}]}}
\newcommand{\RAAYnS}[2][]{{\RAA[#1]\big(\YnS{#2}\big)}}

\newcommand{\Npart}{{N_\text{part}}}
\newcommand{\pT}{{p_\text{T}}}
\newcommand{\sNN}{\sqrt{s_\text{NN}}}

\newcommand{\Teff}{{T_\text{eff}}}
\newcommand{\TDoppler}{{T_\text{D}}}
\newcommand{\tauFnl}{{\tau_{\text{F},nl}}}
\newcommand{\vQGP}{\bm{v}_\text{QGP}}
\newcommand{\AngularAverage}[1]{\langle{#1}\rangle_\Omega}
\newcommand{\AngularMaximum}[1]{{\max}_\Omega({#1})}

\newcommand{\WithUnit}[2]{{#1}\,\textnormal{#2}} % (physical) quantity with unit
\newcommand{\To}{\textnormal{\,--\,}} % delimiter for range of numbers

\usepackage{xcolor}

\newcommand{\note}[2][]{\rlap{\textcolor{red}{\kern-0.1em*}}\marginpar{\raggedright\linespread{1}\scriptsize\textcolor{red}{#2\textsuperscript{\textbf{#1}}}}}

\begin{document}
\preprint{APS/123-QED}

\title{In-medium $\Y$~suppression and feed-down in UU and PbPb~collisions}% Force line breaks with \\
%\thanks{A footnote to the article title}%
\author{J.~Hoelck}
\author{F.~Nendzig}
\author{G.~Wolschin}%
 \email{g.wolschin@thphys.uni-heidelberg.de}
\affiliation{%
 Institut f{\"ur} Theoretische 
Physik
der Universit{\"a}t Heidelberg, Philosophenweg 16, D-69120 Heidelberg, Germany, EU\\
 %Authors' institution and/or address\\
% This line break forced with \textbackslash\textbackslash
}%

%\collaboration{MUSO Collaboration}%\noaffiliation

\date{\today}% It is always \today, today,
             %  but any date may be explicitly specified

\begin{abstract}
The suppression of $\Y$~mesons in the hot quark-gluon medium (QGP) versus reduced feed-down is investigated in UU~collisions at RHIC energies and PbPb~collisions at LHC energies.
Our centrality- and $\pT$-dependent model encompasses screening, collisional damping and gluodissociation in the QGP.
For $\YnS{1}$ it is in agreement with both STAR and CMS data provided the relativistic Doppler effect and the reduced feed-down from the $\YnS{n}$ and $\chibnP{n}$ states are properly considered.
At both energies, most of the $\YnS{1}$ suppression is found to be due to reduced feed-down, whereas most of the $\YnS{2}$ suppression is caused by hot-medium effects. The importance of the latter increases with energy.
The $\pT$-dependence is flat due to the relativistic Doppler effect and reduced feed-down.
We predict the $\YnS{1}$-suppression in PbPb at $\sNN = \WithUnit{5.02}{TeV}$.

\end{abstract}

% PACS, the Physics and Astronomy
                             % Classification Scheme.
%\keywords{Suggested keywords}%Use showkeys class option if keyword
                              %display desired
\maketitle
%%%%%%%%%% BEGIN COPY AND PASTE %%%%%%%%%%%%

\section{Introduction}
\label{intro}
%% main text
%Whereas $\Jpsi$-suppression in relativistic heavy-ion collisions has been investigated in detail for almost 30 years now, the research on
The suppression of bottomia in the hot quark-gluon medium that is created in high-energy heavy-ion collisions at the Relativistic Heavy-Ion Collider (RHIC) and the Large Hadron Collider (LHC) is a new field of research.
The first heavy-ion data were taken by CMS in PbPb~collisions at $\WithUnit{2.76}{TeV}$ center-of-mass energy per particle pair where the $\YnS{1}$, $\YnS{2}$ and $\YnS{3}$ states could be resolved~\cite{CMS-2012}, followed by the ALICE collaboration~\cite{ab14}, 
%-- where the $\YnS{2}$ and $\YnS{3}$ states still wait for their resolution -- 
as well as STAR measurements of $\WithUnit{200}{GeV}$~AuAu~\cite{ada14} and $\WithUnit{193}{GeV}$~UU~\cite{ver16} at RHIC.

The production of heavy mesons and, in particular, of bottomia in initial hard partonic interactions in relativistic heavy-ion collisions at RHIC and LHC energies is of special interest because quarkonia in the fireball can act as a probe to test the properties of the hot medium.
The heavier the hadron that is produced in the collision, the shorter its formation time $\tau_\text{F}$. Very heavy mesons such as the $\Jpsi$ or the $\Y$~meson
%with a rest mass of about $\WithUnit{9.46}{GeV/$c^2$}$
in their $1\mathrm{S}$ spin-triplet ground states are produced in hard collisions at very short times, typically at $\tau_\text{F}\simeq\WithUnit{0.3\To0.5}{fm/$c$}$.
Since the $\YnS{1}$ state is particularly stable, it has a sizeable probability to survive in the hot quark-gluon medium that is produced in the fireball of a heavy-ion collision at LHC energies, even at initial medium temperatures of the order of $\WithUnit{400}{MeV}$ or above.

There exists meanwhile a considerable literature on the dissociation of quarkonia and in particular of the $\Y$~meson in the hot quark-gluon medium; see~\cite{an16} and references therein for a review.
In~\cite{nendzig-wolschin-2013,ngw14} we have devised a model that accounts for the gluon-induced dissociation of the various bottomium states in the hot medium (gluodissociation), the damping of the quark-antiquark binding due to the presence of the medium which generates an imaginary part of the temperature-dependent potential, and the screening of the real part of the potential.
The latter turns out to be less important for the strongly bound $\YnS{1}$ ground state, but it is relevant for the excited $\bb$ states, and also for all $\cc$ bound states.

In this work we utilize our model to quantitatively disentangle the role of bottomia suppression in the quark-gluon plasma (QGP) relative to the role of reduced feed-down for the $\YnS{1}$ ground state as function of energy, comparing with data from both RHIC and LHC. The $\pT$-dependence and the role of the relativistic Doppler effect on the measured transverse-momentum spectra is discussed in detail.
We simultaneously consider the $\YnS{2}$ state where the QGP effects are expected to be much more important with respect to reduced feed-down regarding the measured suppression, and verify this expectation in a calculation.
We compare with centrality-dependent STAR data for UU at $\WithUnit{193}{GeV}$~\cite{ver16} for the $\YnS{1}$ state and CMS data~\cite{CMS-2012,jo15} for the $\YnS{1}$ and $\YnS{2}$ states in $\WithUnit{2.76}{TeV}$ PbPb collisions.
We also predict the centrality-dependent suppression at the higher LHC energy of $\sNN = \WithUnit{5.02}{TeV}$.

We do not include an explicit treatment of cold nuclear matter (CNM) effects in our present study. These are certainly very relevant in asymmetric collisions such as pPb where most of the system remains cold during the interaction time. In symmetric systems at RHIC and LHC energies, however, the CNM effects such as shadowing are likely less important and moreover, expected to be very similar for ground and excited states.
Statistical recombination of the heavy quarks following bottomia dissociation is disregarded as well: Although this is certainly a relevant process in the $\Jpsi$ case, the significantly smaller cross section for $\Y$ production allows us to neglect it.

In the next section, we first describe the hydrodynamic evolution of the hot fireball including transverse expansion. This serves as a simple model for the bulk evolution
since the more specific conclusions of this work regarding the relative importance of in-medium suppression versus reduced feed-down are not expected to depend much
on details of the background model. In section III we outline our phenomenological determination of the initial bottomia populations from an inverse cascade calculation based on
the measured final dimuon yields in pp collisions, scaled by the number of binary collisions. 

Section IV concerns the main part of this work, namely, in-medium dissociation processes as opposed to the effect of
 reduced feed-down. In section V we come to another central part of this manuscript, which is the detailed consideration of the relativistic Doppler effect
due to the relative velocity of the bottomia with respect to the expanding medium. It is shown that this effect -- in combination with the feed-down cascade -- causes the flat transverse momentum dependence of
the suppression factors $\RAA$ that has been observed in recent data \cite{kim16}. The comparison with centrality-dependent data on bottomia suppression at energies obtained at RHIC and at  LHC is presented in section VI, again with emphasis on the role of in-medium effects versus reduced feed-down depending on the state and the incident energy.
The conclusions are drawn in section VII.

%\begin{changes}
\section{Hydrodynamic expansion}
The bottomia states are produced with a formation time $\tauFnl$ in initial hard collisions at finite transverse momentum $\pT$ and then move in the hot expanding fireball where the dissociation processes take place, resulting in a local time dependent decay width $\Gamma_{nl}$ for the states with main quantum number $n$ and angular-momentum quantum number $l$. Before treating their dissociation in detail, we first consider the hydrodynamic flow of the fireball -- the background bulk evolution --, which is basically as outlined in~\cite{ngw14} including transverse expansion.

We describe the QGP by a relativistic, perfect fluid consisting of gluons and massless up-, down- and strange-quarks, whose energy-momentum tensor reads
\begin{equation}
 \mathcal{T} = (\varepsilon + P) u \otimes u + P,	\label{perfect-fluid-em-tensor}
\end{equation}
where $\varepsilon$ is the fluid's internal energy density, $P$ the pressure and $u$ the fluid four-velocity.
For a general energy-momentum tensor the equations of motion are obtained by imposing four-momentum conservation, $\nabla \cdot \mathcal{T} = 0$, which yields
\begin{equation}
 \frac{1}{\sqrt{|\det g|}} \partial_\mu \left( \sqrt{|\det g|} \mathcal{T}^\mu{}_\alpha \right) = \frac{1}{2} \mathcal{T}^{\mu\nu} \partial_\alpha g_{\mu\nu}\,,
 \label{energy-momentum-conservation}
\end{equation}
where $g = g_{\mu\nu} \mathrm{d}x^\mu \mathrm{d}x^\nu$ is the spacetime-metric and Eq.~(\ref{perfect-fluid-em-tensor}) has to be inserted for $\mathcal{T}$. The system of equations is closed by the equation of state, appropriate for a perfect, relativistic fluid,
\begin{equation}
 P = c_\text{s}^2 \varepsilon,\qquad		c_\text{s} = \frac{1}{\sqrt{3}},\qquad		\varepsilon = \varepsilon_0 T^4.	\label{EOS}
\end{equation}
We evaluate Eq.~(\ref{energy-momentum-conservation}) in the longitudinally co-moving frame (LCF), where the metric $g$ is given by
\begin{equation}
 g = -\mathrm{d}\tau^2 + \tau^2 \mathrm{d}y^2 + (\mathrm{d}x^1)^2 + (\mathrm{d}x^2)^2,	\label{metric}
\end{equation}
with the $x^1$-axis lying within and the $x^2$-axis orthogonal to the reaction plane.
In this frame the fluid flour-velocity $u$ reads
\begin{gather}\label{fluid-velocity}
 u = \gamma_\perp (e_\tau + v^1 e_1 + v^2 e_2)~,\\
 \gamma_\perp = \frac{1}{\sqrt{1 - (v^1)^2 - (v^2)^2}}~.
\end{gather}
Note that the same transverse velocity components $v^1$, $v^2$ are measured in the laboratory frame (LF) as in the LCF; a property that is very convenient when dealing with quantities that depend on transverse momentum $\pT$. Inserting Eqs.~(\ref{perfect-fluid-em-tensor}) and (\ref{EOS})\To(\ref{fluid-velocity}) into Eq.~(\ref{energy-momentum-conservation}) yields
\begin{equation}
 \partial_\mu (\tau T^4 u^\mu u_\alpha) = - \frac{\tau}{4} \partial_\alpha T^4,	\qquad
 \partial_\mu (\tau \, T^3 u^\mu) = 0,	\label{EOM}
\end{equation}
where the second equation corresponds to $u \cdot (\nabla \cdot \mathcal{T}) = 0$.

We solve Eqs.~(\ref{EOM}) numerically, starting at the initial time $\tau_\text{init} = \WithUnit{0.1}{fm/$c$}$ in the LCF.
The initial conditions in the transverse plane $(x^1,x^2)$ are given in Eqs.\,(14)\To(16) of~\cite{ngw14} as
\begin{gather}
 v^1(\tau_{\text{init}}) = v^2(\tau_{\text{init}}) = 0 \\
 T(b, \tau_{\text{init}}, x^1, x^2) = T_0 \left( \frac{N_\text{mix}(b,x^1,x^2)}{N_\text{mix}(0,0,0)} \right)^{1/3} \\
 N_{\text{mix}}^{\text{RHIC}} = \frac{1 - f}{2} N_{\text{part}} + f N_{\text{coll}},\quad f = 0.145 \\
 N_{\text{mix}}^{\text{LHC}} = \hat{f} N_{\text{part}} + (1-\hat{f}) N_{\text{coll}},\quad \hat{f} = 0.8
\end{gather}
where $f,\hat{f}$ are from~\cite{PHOBOS-2004,ALICE-2011a} and $b$ is the impact parameter.
%,$T_0 = \WithUnit{480}{MeV}$ for PbPb 
% (see~\cite{nendzig-phd} for more information on the numerical procedure). Temperature profiles are shown in figure \ref{fig:profile-T} for a central collision ($b = 0$).
The initial central temperature $T_0$ is fixed through a fit of the $\pT$-dependent minimum-bias experimental $\RAAYnS{1}$ results for PbPb at $\WithUnit{2.76}{TeV}$, cf.~section~\ref{doppler-effect}.
For other systems and incident energies, $T_0$ is scaled consistently with respect to the produced charged hadrons.
%It is in reasonable agreement with values derived from ideal hydrodynamical calculations of elliptic flow at RHIC and LHC energies, respectively~\cite{schen11}.

We define the QGP-suppression factors $R^\text{QGP}_{AA,nl}(c,\pT)$ which quantify the amount of in-medium suppression of bottomia with transverse momentum $\pT$ for PbPb collisions in the centrality bin $c$, where $b_c \leq b < b_{c+1}$. The QGP-suppression factor is not directly measurable since it accounts only for the amount of suppression inside the fireball due to the three processes of color screening, collisional damping and gluodissociation that we consider in the next sections. It is given by the ratio of the number of bottomia that have survived the fireball to the number of bottomia produced in the collision. The latter scales with the number of binary collisions at a given point in the transverse plane and hence with the nuclear overlap, $N_\bb \propto N_\text{coll} \propto T_{AA}$. Thus we write $R_{AA,nl}^\text{QGP}$ as follows:
\begin{multline}\label{RAAQGP}
 R_{AA,nl}^\text{QGP}(c,\pT)\\
 =\frac{\int_{b_c}^{b_{c+1}} \mathrm{d}b \, b \int \mathrm{d}^2x \, T_{AA}(b,x^1,x^2) \, D_{nl}(b,\pT,x^1,x^2)}{\int_{b_c}^{b_{c+1}} \mathrm{d}b \, b \int \mathrm{d}^2x \, T_{AA}(b,x^1,x^2)}~.
\end{multline}
The damping factor $D_{nl}$ is determined by the temporal integral over the corresponding $\bb$ decay width $\Gamma_{nl}$,
\begin{multline}\label{damping-factor}
 D_{nl}(b,\pT,x^1,x^2)\\
 = \exp\left[ - \int\limits_{\tau_{\text{F},nl} \gamma_{\text{T},nl}(\pT)}^\infty \frac{\mathrm{d}\tau \, \Gamma_{nl}(b,\pT,\tau,x^1,x^2)}{\gamma_{\text{T},nl}(\pT)}\right],
\end{multline}
%\note[GW]{Variable in Gamma einfuegen?}
where $\tau_{\text{F},nl}$ is the formation time in the bottomium rest-frame, $\gamma_{\text{T},nl}(\pT) = \sqrt{1 + (\pT/M^{\text{vac}}_{nl})^2}$ the Lorentz-factor due to transverse motion in the LCF, and $M^{\text{vac}}_{nl}$ the experimentally measured bottomium vacuum mass.
%In particular, for a bottomium state formed initially at the location $(x^1,x^2)$ that moves through the medium with transverse velocity $\beta_{nl}(\pT)$, the exact functional form of the total decay width reads
%\begin{equation}
% \Gamma^\text{tot}_{nl} = \Gamma^\text{tot}_{nl}\big(T_{\text{eff},nl}(b,\pT,\tau,x^1 + \beta^1_{nl} \tau,x^2 + \beta^2_{nl} \tau)\big).
%\end{equation}
\section{Initial bottomia populations}
To estimate the initial populations $N^\text{i}_{AA,nl}$ of the six bottomia states that we treat explicitly in this work, we consider the measured final populations $N^\text{f}_{\text{pp},nl}$ of the three $\YnS{n}$-states in pp collisions at the same energy and calculate the decay cascade~\cite{vaccaro-etal-2013} backwards to obtain the initial populations in pp, $N^\text{i}_{\text{pp},nl}$, shown in Tab.~\ref{tab-population}.
These are then scaled by the number of binary collisions $N_\text{coll}$ yielding the initial populations in the heavy-ion case. 
%$N^\text{i}_{AA,nl} = N_\text{coll} N^\text{i}_{\text{pp},nl}$. 
When the suppression factors are calculated, the number of binary collisions cancels out.
The required branching ratios are taken from the Review of Particle Physics~\cite{PDG-2014} or from theory where no experimental values are available (as is the case for 
$\chibnP{3}$), see~\cite{vaccaro-etal-2013} for details and references.
%\end{changes}

\begin{table}
	\caption{\label{tab-population}
		Initial populations of the different bottomium states as obtained from an inverted feed-down cascade calculation in pp collisions at $\WithUnit{2.76}{TeV}$ normalized by the $\YnS{1}$ population after feed-down, $n^\text{i}_{\text{pp},nl} = N^\text{i}_{\text{pp},nl}/N^\text{f}_{\text{pp},\YnS{1}}$.
		($\Gamma_\chibnP{3}$ denotes the yet unkown vacuum decay width of the $\chibnP{3}$ state which cancels out in the computation of final populations.)
	}
	\bigskip
	\centering
	\begin{tabular}{rr}
		\hline\hline
		State & $n^\text{i}_{\text{pp},nl}$ \\
		\hline
		$\YnS{1}$ & $0.373$ \\
		$\chibnP{1}$ & $1.084$ \\
		$\YnS{2}$ & $0.367$ \\
		$\chibnP{2}$ & $0.881$ \\
		$\YnS{3}$ & $0.324$ \\
		$\chibnP{3}$ & $0.00835\,\text{eV}^{-1} \Gamma_\chibnP{3}$ \\
		\hline\hline
	\end{tabular}
\end{table}
Regarding the production process, we use the same formation time of $\tauFnl = \WithUnit{0.4}{fm/$c$}$ for ground and exited states, with theta functions for the production as function of time.
In the co-moving coordinate system used for the hydrodynamical calculation, time dilation of the formation times is then taken into account.
As has been indicated e.g. in \cite{ko15}, the quarkonium formation time in heavy ion collisions is not well determined. We had investigated the dependence of our model results on $\tauFnl$ to some extent in our previous article (Fig.9, Tab. II in \cite{ngw14}) and in the present work we keep it fixed for all states. Clearly this is an idealization, and the medium and temperature effects on $\tauFnl$ need to be investigated further \cite{ko15,gami15}.
%We had previously investigated the dependence of $\RAA$ on $\tauFnl$, cf.~Fig.~9 in~\cite{ngw14}.

%\begin{changes}
\section{In-medium dissociation versus reduced feed-down}
To obtain the wave functions and eventually the decay widths of the bottomia states considered at each space-time point and temperature in the hot fireball, we solve the radial Schr\"odinger equation for the six states $\YnS{1}$, $\YnS{2}$, $\YnS{3}$, and $\chibnP{1}$, $\chibnP{2}$, $\chibnP{3}$~with energies $E_{nl}(T)$ in a complex potential $V_{nl}(r,T)$ \cite{ngw14} and corresponding damping widths $\Gamma^\text{damp}_{nl}(T)$,
\begin{multline}\label{radialschroedinger}
 \partial_r^2 g_{nl}(r,T)\\
 = m_\mathrm{b} \left( V_{nl}(r,T) - E_{nl}(T) + \frac{\mathrm{i}\Gamma^\text{damp}_{nl}(T)}{2} \right) g_{nl}(r,T)\,.
\end{multline}
Here $m_\mathrm{b}$ is the $\mathrm{b}$-quark mass and $T$ the QGP temperature at any given space-time point. 

We consider the running of the strong coupling in the calculation of the wave function, and the various dissociation processes.
In the complex potential $V_{nl}(r,T)$ \cite{ngw14} a variable $\alpha_{nl}$ appears that denotes the strong coupling $\alpha_s$ evaluated at the soft scale $S_{nl}(T) = \langle 1/r\rangle_{nl}(T)$.
Hence $\alpha_{nl}$ depends on the solution $g_{nl}$ of the Schr\"odinger equation for the six bottomium states through the averaging, and we use an iterative method for the solution of the problem~\cite{ngw14}, together with the one-loop expression for the running of the coupling.

In addition to the damping width $\Gamma^\text{damp}_{nl}(T)$ we calculate the width caused by gluon-induced dissociation $\Gamma^\text{diss}_{nl}(T)$~\cite{brezinski-wolschin-2012,ngw14}.
The total in-medium decay width of a given bottomium state is the incoherent sum $\Gamma^\text{tot}_{nl} = \Gamma^\text{damp}_{nl} + \Gamma^\text{diss}_{nl}$.
The two mechanism emerge in different orders in the effective action, as has been shown in potential nonrelativistic~QCD approaches~\cite{bram08,brambilla-etal-2011}.
The imaginary part of the interaction potential $V_{nl}$ yields collisional damping (``soft process'' in pNRQCD terminology) whereas gluodissociation is described by a singlet to octet transition (``ultrasoft process''), and hence both should be treated individually due to the separation of scales.

In~\cite{brezinski-wolschin-2012} we had derived the gluodissociation cross section for a screened Cornell-type potential with temperature-dependent string part through an extension of the operator product expansion that was developed in~\cite{bhanot-peskin-1979} for Coulomb-like momentum eigenstates.
The result agrees with the one obtained independently in effective field theory~\cite{brambilla-etal-2011} in the corresponding limit.

Once the bottomia states have survived the hot quark-gluon plasma environment, the feed-down cascade from the excited states to the ground state is considered in detail.
Due to the rapid melting or depopulation of the excited states caused by the mechanisms in the QGP-phase, the feed-down to the ground state is reduced, resulting in additional $\YnS{1}$-suppression with respect to the situation in pp~collisions at the same energy.
%other spin-triplet states are above the $\mathrm{B\bar{B}}$~threshold and hence, are not relevant for a comparison with the measured $\YnS{n}$-suppression.
%The transverse-momentum dependence of the suppression is discussed for minimum-bias collisions (averaged over all centralities, $\WithUnit{0\To100}{\%}$).

The focus of the present investigation is the determination of the relevance of reduced feed-down for a given bottomium state as function of incident energy (RHIC vs. LHC), and of its relative importance for the $\YnS{1}$ and $\YnS{2}$ states, which appears to be a new consideration.
As will be shown in the following sections, at both RHIC and LHC energies, most of the suppression for the $\YnS{1}$ state is found to be due to reduced feed-down, and even more so at the lower RHIC energy.
In contrast, most of the $\YnS{2}$ suppression is caused by the hot-medium effects.

\section{Relativistic Doppler effect}\label{doppler-effect}
% Instead of an isotropic temperature $T$, moving bottomia are expected to be subject to an anisotropic temperature $\TDoppler$ due to the relativistic Doppler effect:
Bottomia are too massive to experience a substantial change of their momenta by collisions with the light partons in the medium.
Hence, there will be a finite relative velocity between the expanding QGP medium and the $\bb$ mesons.
In the rest frame of the bottomia, the surrounding distribution of massless gluons then appears as a Bose-Einstein distribution with an anisotropic temperature $\TDoppler$ that is determined from the relativistic Doppler effect as
\begin{equation}
	\TDoppler(T,|\vQGP|,\Omega) = T \frac{\sqrt{1 - |\vQGP|^2}}{1 - |\vQGP| \cos\theta}\,.
	\label{TDoppler}
\end{equation}
Here, $\vQGP$ is the average velocity of the surrounding fluid cell (measured in the bottomium restframe) and $\Omega = (\theta,\phi)$ the solid angle where $\theta$ measures the angle between $\vQGP$ and the incident light parton.
%direction in which the temperature is evaluated.

%This formulation is valid for gluodissociation and damping, e.\,g.~the imaginary part of the potential. Here we extend it to the modification of the screening of the real part of the %potential: It is thus taken into account also for the evolution of screened bound states that are moving at finite velocity.
%\begin{changes}
%as was also concluded in~\cite{escobedo-etal-2011}.
In the rest frame of the bottomia, the Doppler effect causes a blueshifted temperature for $\theta = 0^\circ$ and a redshifted temperature in the opposite direction $\theta = 180^\circ$.
%\end{changes}
The effects of red- and blueshift get more and more pronounced with increasing relative velocity~$|\vQGP|$, but the angular range with $\TDoppler < T$ (redshifted region) is growing while the angular range with $\TDoppler > T$ (blueshifted region) is restricted to smaller and smaller angles~$\theta$ as has been noted in~\cite{escobedo13}. 
To account for the effect of the anisotropic temperature $\TDoppler$ on the bottomium dissociation, we explore different possibilities.

First, one can estimate the upper limit of the impact of the Doppler shift on the suppression factor~$\RAA$ by considering only the maximum blueshift, substituting $T$ by an effective temperature
\begin{equation}
	\Teff \coloneqq \AngularMaximum{\TDoppler}\,.
\end{equation}
%We find at small but finite $\pT$ (up to $\WithUnit{2}{GeV/$c$}$) less suppression compared to the case $\vQGP=\bm{0}$ because the %bottomia tend to escape the QGP zone faster, but in a $\pT$-range of $\WithUnit{2\To10}{GeV/$c$}$ the blueshifted temperature prevails %and causes more suppression, see Fig.~\ref{fig1}.
%Indeed the preliminary minimum-bias CMS data~\cite{jo15} show no significant $\pT$-dependence. 
As expected, we find substantial suppression in the whole $\pT$ range, more than what is seen in the CMS data~\cite{jo15} for 
$\WithUnit{4}{GeV/$c$} < \pT < \WithUnit{16}{GeV/$c$}$.

\begin{figure}
	\centering
	\includegraphics[scale=0.8]{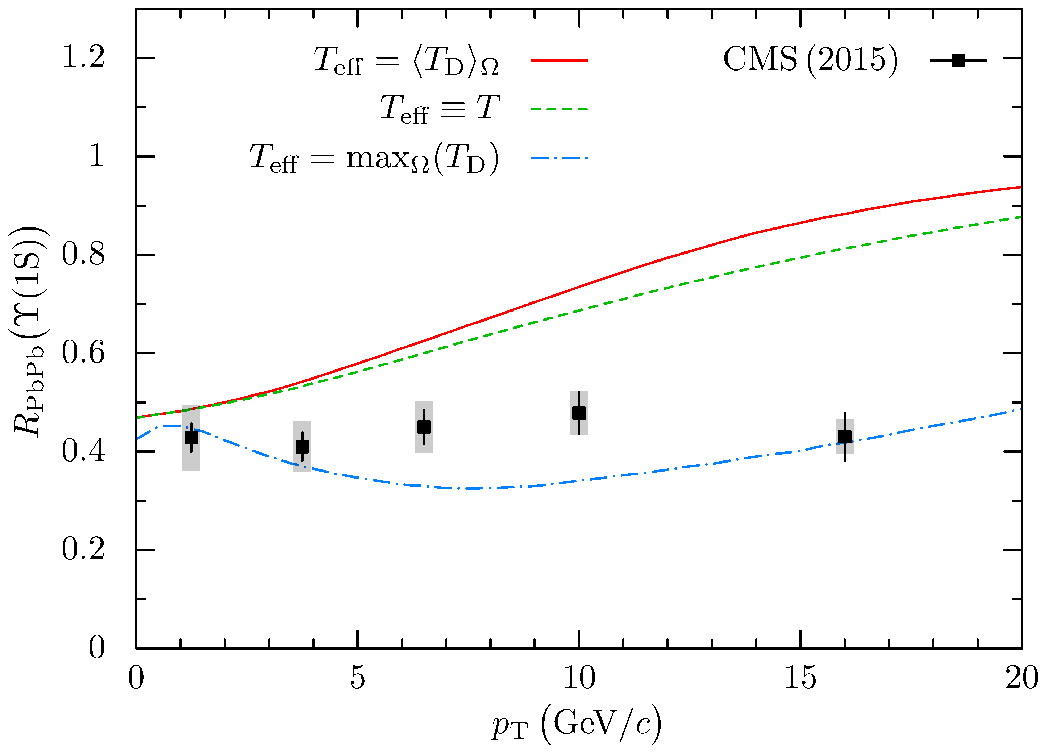}
	\caption{\label{fig1}
		(color online)
		Transverse-momentum dependence of the ground-state suppression factor $\RAAYnS{1}$ in minimum-bias PbPb collisions at $\sNN = \WithUnit{2.76}{TeV}$ for effective temperatures based on maximum blueshift only $\Teff = \AngularMaximum{\TDoppler}$ (dash-dotted, lower curve) and angular-averaged temperature $\Teff = \AngularAverage{\TDoppler}$ (solid, upper curve) in comparison with the unmodified case $\Teff \equiv T$ (dashed, middle curve). Preliminary data from CMS~\cite{jo15} at $\WithUnit{2.76}{TeV}$ are shown for comparison. Statistical error bars are solid, systematic ones shaded.
	}
\end{figure}

A second possibility to approximate the effect of the anisotropic temperature is to use the angular average of $\TDoppler$ as effective temperature,
\begin{equation}
	\Teff \coloneqq \AngularAverage{\TDoppler} = \frac{1}{4\pi}\int\mathrm{d}\Omega~\TDoppler\,.
\end{equation}
As shown in Fig.~\ref{fig1}, in this case the impact of the redshift outbalances the blueshift for all relative velocities and $\RAA$ rises monotonically with $\pT$ as already discussed in~\cite{ngw14}. The data~\cite{jo15} fall in between the two cases of maximum blueshift only and angular-averaged temperature.

The choice $\Teff = \max_\Omega(\TDoppler)$ systematically overestimates the effective temperature because it ignores any redshifted contributions.
Evaluating the decay widths at an angular-averaged $\TDoppler$ avoids this shortcoming but 
leads to unphysical results when the temperature in the blueshifted region exceeds the dissociation temperature:
In this case the existence of bound states should be prohibited, corresponding to an infinitely large decay width, but the averaging can artificially lower the temperature to a value where bound states can exist.

A better approach is provided by directly substituting the total decay width $\Gamma^\text{tot}_{nl}(T)$ by an effective, angular-averaged value
%averaging the total decay width $\Gamma^\text{tot}_{nl}$ over $\Omega$,
\begin{equation}
	\Gamma^\text{tot}_{\text{eff},nl} \coloneqq \AngularAverage{\Gamma^\text{tot}_{nl}(\TDoppler)} = \frac{1}{4\pi}\int\mathrm{d}\Omega~\Gamma^\text{tot}_{nl}(\TDoppler)\,.
	\label{eqaverageing}
\end{equation}
This takes into account the redshifted temperatures and also correctly describes the non-existence of bound states once the effective temperature in the blueshifted region exceeds the dissociation temperature.

\begin{figure}
	\centering
	\includegraphics[scale=0.8]{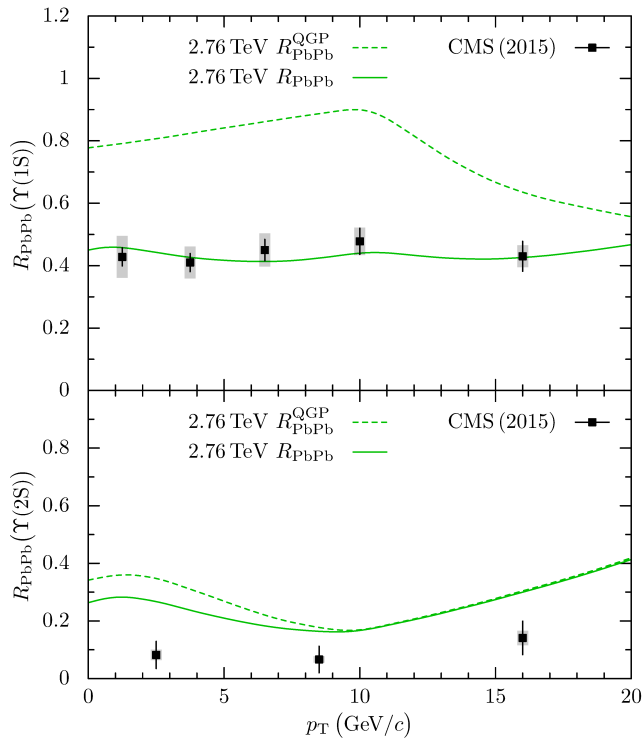}
	\caption{\label{fig2}
		(color online)
		Transverse-momentum dependence of the suppression factors $\RAAYnS{1}$ for the ground state and $\RAAYnS{2}$
		for the first excited state  in minimum-bias PbPb~collisions at $\sNN = \WithUnit{2.76}{TeV}$. The (upper) dashed curves show the suppression in the hot medium, the (lower) solid curves the suppression including reduced feed-down, which is only important for the ground state.
		% and $\WithUnit{5.02}{TeV}$ (dashed curve) calculated according to Eq.\,(\ref{eqaverageing}). Preliminary CMS data are from~\cite{jo15}.
	}
\end{figure}
%The Doppler effect can raise low temperatures artificially above the QGP formation temperature $\Tcrit$ although no quark-gluon medium is present, which would distort the physical values of the widths. We have coupled the existence of the QGP explicitly to the unshifted medium temperature $T$ rather than the effective temperature $T_\mathrm{eff}$ to prevent this.

The resulting $\pT$-dependence of the $\YnS{1}$ and $\YnS{2}$ suppression is shown in Fig.~\ref{fig2} for minimum-bias PbPb~collisions at $\WithUnit{2.76}{TeV}$ together with the preliminary CMS data~\cite{jo15}. 
%and our prediction at $\WithUnit{5.02}{TeV}$~PbPb. 
The in-medium suppression factor $\RAAQGP$ for the ground state rises towards $\pT \simeq \WithUnit{10}{GeV/$c$}$ because with increasing $\pT$ it becomes easier for the Upsilon to leave the hot zone.
At $\WithUnit{10}{GeV/$c$} \lesssim \pT \lesssim \WithUnit{20}{GeV/$c$}$ the rising widths overcompensate this trend, causing a fall of $\RAAQGP$.
When the reduced feed-down from the excited states is considered, the suppression factor $\RAA$ becomes rather flat, in reasonable agreement with the available CMS data for the ground state. 

For the $\YnS{2}$ state the calculated $\pT$-dependence reproduces the trend seen in the data, but we underestimate the suppression, evidently because additional mechanisms are at work that we have not yet considered.
The local maximum is here at considerably lower values of $\pT$ than for the ground state, followed by a local minimum and a steady rise. Reduced feed-down has a small effect only at $\pT$ below $\WithUnit{10}{GeV/$c$}$, it is much less important than for the ground state.

\begin{figure}
	\centering
	\includegraphics[scale=0.8]{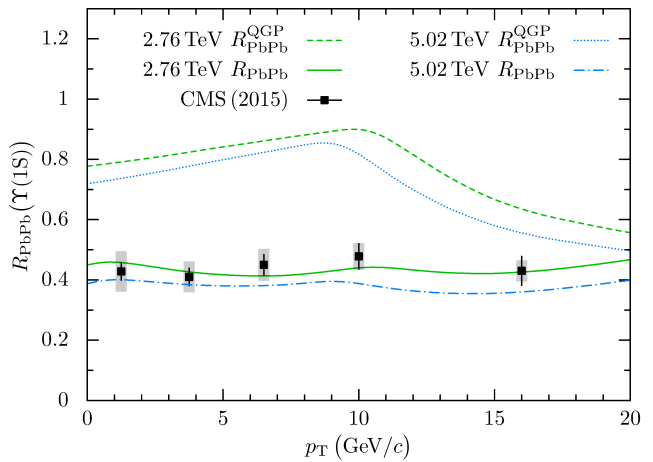}
	\caption{\label{fig3}
		(color online)
		Transverse-momentum dependence of the suppression factors $\RAAQGP$ (dashed line) and $\RAA$ (solid line) for the ground state in minimum-bias PbPb~collisions at $\sNN = \WithUnit{2.76}{TeV}$ ($T_0 = \WithUnit{480}{MeV}$) compared with recent CMS data~\cite{jo15}.
		Our predictions for $\WithUnit{5.02}{TeV}$~PbPb ($T_0 = \WithUnit{513}{MeV}$) are shown as dotted and dash-dotted curves for $\RAAQGP$ and $\RAA$, respectively.
		% and $\WithUnit{5.02}{TeV}$ (dashed curve) calculated according to Eq.\,(\ref{eqaverageing}). Preliminary CMS data are from~\cite{jo15}.
	}
\end{figure}

The predicted energy dependence of the ground-state suppression is shown in Fig.~\ref{fig3}: We find slightly more suppression, 
however, compatible with the 2.76 TeV result within the experimental error bars.
%In both cases the $\pT$-dependence is rather flat and at $\WithUnit{2.76}{TeV}$ in reasonable agreement with the available data. 
%Hence the formulation based on the relativistic Doppler shift with angular-averaged decay widths is consistent with the flat $\pT$-dependence %seen in the data for the $\YnS{1}$ suppression provided the reduced feed-down is properly taken into account.
%\note[JH]{L\"oschen?}
%In our previous approach~\cite{ngw14} we had used the case of angular-averaged effective temperature that produces predominantly redshift and hence, insufficient suppression at %large transverse momentum. 

The centrality dependence that we obtain after averaging over $\pT$ is not very different compared to our previous results \cite{ngw14} at 2.76 TeV calculated with an angular-averaged effective temperature.
There is presently no rapidity dependence in our model, both minimum bias and centrality dependent yields are flat as functions of $y$, corresponding to a boost invariant hydrodynamical evolution.
%\begin{table}
%	\caption{\label{tab-decaywidths}
%		Effective total decay widths for $T = \WithUnit{170}{MeV}$ and various medium velocities $|\vQGP|$ in the bottomium restframe.
%	}
%	\bigskip
%	\centering
%	\begin{tabular}{rrrrr}
%	%\vspace{0.2cm}
%		\hline\hline\\
%		%\vspace{0.3cm}
%		& \multicolumn{4}{c}{$\AngularAverage{\Gamma^\text{tot}_{nl}(\TDoppler)}$~(MeV)} \\
%		State & $|\vQGP|=0$ & $|\vQGP|=0.2$ & $|\vQGP|=0.4$ & $|\vQGP|=0.6$ \\
%		\hline\\
%		$\YnS{1}$ & $35.15$ & $35.79$ & $37.64$ & $41.1$ \\
%		$\chibnP{1}$ & $166.7$ & $169.3$ & $178.0$ & --- \\
%		$\YnS{2}$ & $212.7$ & $213.1$ & $213.5$ & --- \\
%		% $\chibnP{2}$
%		% $\YnS{3}$
%		% $\chibnP{3}$
%		\hline\hline
%	\end{tabular}
%\end{table}
%For comparison with other approaches we display angular-averaged total decay widths of bottomia states at temperature $T = \WithUnit{170}{MeV}$ and different medium velocities $|\vQGP|$ in Tab.~\ref{tab-decaywidths}. At this temperature and above, the total in-medium widths become rather flat as functions of $|\vQGP|$.
%At very high velocities, the excited states are screened due to the high effective temperatures, as shown here for the $\chibnP{1}$ and $\YnS{2}$ states.

\section{Centrality-dependent results and comparison to data}\label{results}
\begin{figure}
	\centering
	\includegraphics[scale=0.8]{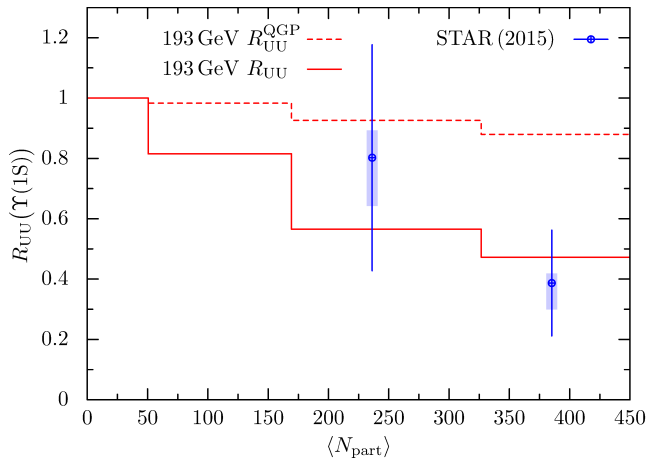}
	\caption{\label{fig4}
		(color online)
		Calculated suppression factor for the $\Y$~spin-triplet ground state $\RAAYnS{1}$ in UU~collisions at $\sNN = \WithUnit{193}{GeV}$ (solid line) together with centrality-dependent preliminary data ($\WithUnit{10\To30}{\%}$ and $\WithUnit{0\To10}{\%}$, $|y| < 1$, circles) from STAR~\cite{ver16}
		as function of the number of participants $\Npart$ (averaged over centrality bins). The suppression factor $\RAAQGP$ in the QGP-phase without the effect of reduced feed-down is shown as dashed (upper) curve.
		%The largest fraction of the suppression is due to reduced feed-down.
		%A prediction of $\RAAYnS{1}$ for $\sNN = \WithUnit{5.02}{TeV}$ (dash-dotted line) as a function of centrality, averaged over $\pT$, is plotted for a formation time $\tauFnl = \WithUnit{0.4}{fm/$c$}$, lowest curve. Here the initial central temperature $T_0 = \WithUnit{610}{MeV}$ is obtained from scaling according to the initial entropy density.
	}
\end{figure}

\begin{figure}
	\centering
	\includegraphics[scale=0.8]{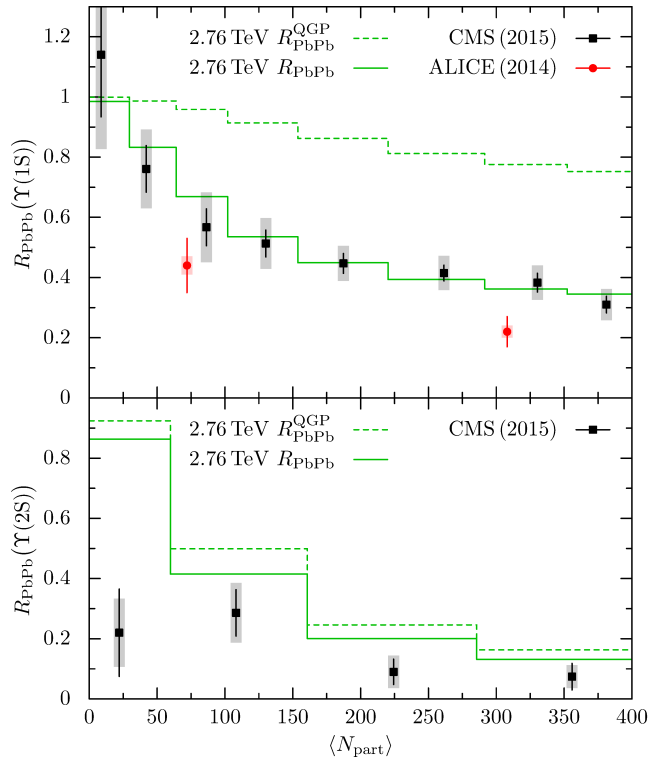}
	\caption{\label{fig5}
		(color online)
		Top: Calculated suppression factor $\RAAYnS{1}$ in PbPb~collisions at $\sNN = \WithUnit{2.76}{TeV}$ (solid line) together with centrality-dependent data from CMS (squares, $|y| < 2.4$,~\cite{jo15}) and ALICE (circles, $2.5 < y < 4$,~\cite{ab14}) as function of the number of participants $\Npart$ (averaged over centrality bins).
		The suppression factor $\RAAQGP$ in the QGP-phase without the effect of reduced feed-down is shown as dashed (upper) curve.
		%A prediction of $\RAAYnS{1}$ for $\sNN = \WithUnit{5.02}{TeV}$ (dash-dotted line, and dotted line for corresponding $\RAAQGP$) as a function of centrality is plotted for %a formation time $\tauFnl = \WithUnit{0.4}{fm/$c$}$.
		Bottom: Suppression factor for the first excited state $\RAAYnS{2}$ in PbPb~collisions at $\sNN = \WithUnit{2.76}{TeV}$ (solid line) together with preliminary data from CMS~\cite{jo15}. The suppression factor $\RAAQGP$ in the QGP-phase (dashed) accounts for most of the calculated total suppression (solid) for the $\YnS{2}$.
		%Most of the suppression for excited states occurs in the QGP-phase. Additional suppression mechanisms are required in particular for peripheral collisions.
	}
\end{figure}

As is displayed in Figs.~\ref{fig4} and \ref{fig5}, the suppression of the spin-triplet ground state at both RHIC and LHC energies is well described by our model for initial central temperatures of $T_0 = \WithUnit{417}{MeV}$ in UU at $\sNN = \WithUnit{193}{GeV}$ and $T_0 = \WithUnit{480}{MeV}$ in PbPb at $\WithUnit{2.76}{TeV}$, and an $\Y$ and $\chib$ formation time of $\tauFnl = \WithUnit{0.4}{fm/$c$}$.
% or equivalently $T_0 = \WithUnit{540}{MeV}$ and $\tauFnl = \WithUnit{0.3}{fm/$c$}$.
The parameters for the density distributions of the lead and uranium ions are taken from~\cite{vries87}.
Our minimum bias value of the suppression in $\WithUnit{2.76}{TeV}$~PbPb is $\RAAYnS[\text{min.\,bias}]{1} = 0.43$.

A comparison of Figs.~\ref{fig4} and \ref{fig5} (top) reveals how the relative contributions of in-medium effects and reduced feed-down change as a function of incident energy, see Tab.~\ref{tab3} for detailed minimum-bias results: In $\WithUnit{193}{GeV}$~UU, only about $\WithUnit{20}{\%}$ of the total suppression $(1-\RAA)$ is due to the in-medium effects, whereas in $\WithUnit{2.76}{TeV}$~PbPb the in-medium contribution is already about $\WithUnit{30}{\%}$ and further increases in $\WithUnit{5.02}{TeV}$~PbPb.

The situation is very different for the first excited state $\YnS{2}$ as shown in Fig.~\ref{fig5} (bottom) for $\WithUnit{2.76}{TeV}$~PbPb:
With the same set of parameters as for the $\YnS{1}$ state, the calculated suppression of the $\YnS{2}$ state is much more pronounced in the QGP-phase than for the $\YnS{1}$ state.
Tab.~\ref{tab3} shows that more than $\WithUnit{80}{\%}$ of the total minimum-bias $\YnS{2}$ suppression in UU and more than $\WithUnit{90}{\%}$ in PbPb is due to in-medium effects. Hence the additional contribution of the feed-down cascade to the $\YnS{2}$ suppression is rather marginal and drops below $\WithUnit{10}{\%}$ at LHC energies.
% in particular for central collisions.
Unfortunately, there are no data available that directly quantify the feed-down fractions as functions of centrality.
As is obvious from Fig.~\ref{fig5}, the comparison with the CMS data~\cite{jo15} leaves room for additional suppression mechanisms in particular in the peripheral region for the $\YnS{2}$ state.

\begin{figure}
	\centering
	\includegraphics[scale=0.8]{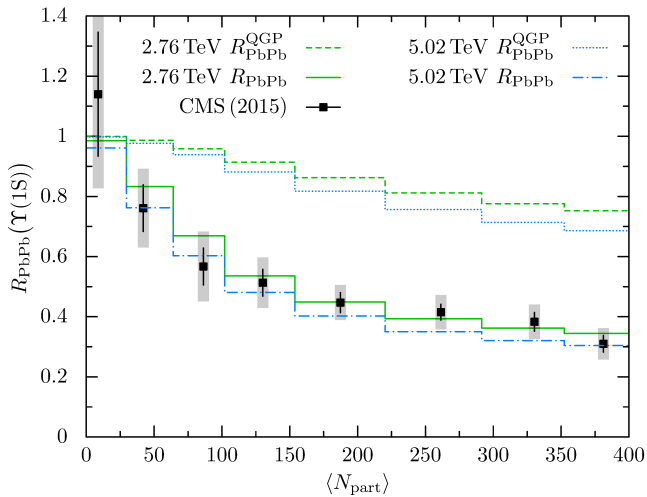}
	\caption{\label{fig6}
		(color online)
		Calculated suppression factor $\RAAYnS{1}$ in PbPb~collisions at $\sNN = \WithUnit{2.76}{TeV}$ (lower solid line) and prediction at $\WithUnit{5.02}{TeV}$ (lower dash-dotted line) together with centrality-dependent $\WithUnit{2.76}{TeV}$ data from CMS (squares, $|y| < 2.4$,~\cite{jo15}) as function of the number of participants $\Npart$ (averaged over centrality bins).
		The suppression factors $\RAAQGP$ in the QGP-phase without the effect of reduced feed-down are shown as upper curves (dashed and dotted) for both energies, again yielding slightly more suppression at the higher energy.
		%A prediction of $\RAAYnS{1}$ for $\sNN = \WithUnit{5.02}{TeV}$ (dash-dotted line, and dotted line for corresponding $\RAAQGP$) as a function of centrality is plotted for %a formation time $\tauFnl = \WithUnit{0.4}{fm/$c$}$.
		% Here the initial central temperature $T_0 = \WithUnit{610}{MeV}$ is obtained from scaling according to the initial entropy density.
		%Bottom: Suppression factor for the first excited state $\RAAYnS{2}$ in PbPb~collisions at $\sNN = \WithUnit{2.76}{TeV}$ (solid line) together with preliminary data from %CMS~\cite{jo15}. The suppression factor $\RAAQGP$ in the QGP-phase without the effect of reduced feed-down is shown as dashed (upper) curve.
		%Most of the suppression for excited states occurs in the QGP-phase. Additional suppression mechanisms are required in particular for peripheral collisions.
	}
\end{figure}

For $\YnS{1}$ we also predict in Fig.~\ref{fig6} the suppression at the higher LHC energy of $\WithUnit{5.02}{TeV}$~PbPb as $\RAAYnS[\text{min.\,bias}]{1} = 0.39$ using the same formation times $\tauFnl = \WithUnit{0.4}{fm/$c$}$, but a scaled initial temperature $T_0 = \WithUnit{513}{MeV}$.
This value is obtained from the proportionality between the initial entropy density, the charged particle multiplicity per unit of rapidity and the cube of the temperature~\cite{bjorken-1983,baym-etal-1983,gyulassi-matsui-1984}.
Here the extrapolated $\mathrm{d}N_\text{ch}/\mathrm{d}\eta$ for $\WithUnit{0\To5}{\%}$~centrality PbPb taken from~\cite{gw15} is in agreement with recent data from ALICE~\cite{adam15}. The ensuing enhancement of the suppression is within the experimental error bars of the $\WithUnit{2.76}{TeV}$ result. It remains to be seen whether this is confirmed by the forthcoming analysis of the mid-rapidity CMS data from LHC Run-II.

\begin{table}
	\centering
	\caption{\label{tab3}
		Calculated nuclear suppression factors for the $\YnS{1}$ and $\YnS{2}$ states in minimum-bias $\WithUnit{193}{GeV}$~UU as well as $2.76$ and $\WithUnit{5.02}{TeV}$~PbPb collisions. The in-medium suppression factor is $\RAAQGP$, the total suppression factor including reduced feed-down is $\RAA$. The last column gives the percentage of the suppression in the medium relative to the total suppression $(1-\RAAQGP)/(1-\RAA)$.
	}
	\bigskip
	\begin{tabular}{rrrrrr}
		\hline\hline
		\vphantom{$\displaystyle\int^0$}
		$\sNN$ & System & State & $\RAAQGP$ & $\RAA$ & $\frac{1-\RAAQGP}{1-\RAA}$\\
		\hline
		$\WithUnit{193}{GeV}$ & UU & $\YnS{1}$ & $0.92$ & $0.57$ & $\WithUnit{19}{\%}$\\
		$\WithUnit{193}{GeV}$ & UU & $\YnS{2}$ & $0.48$ & $0.41$ & $\WithUnit{88}{\%}$\\
		\hline
		$\WithUnit{2.76}{TeV}$ & PbPb & $\YnS{1}$ & $0.83$ & $0.43$ & $\WithUnit{31}{\%}$\\
		$\WithUnit{2.76}{TeV}$ & PbPb & $\YnS{2}$ & $0.28$ & $0.23$ & $\WithUnit{94}{\%}$\\
		\hline
		$\WithUnit{5.02}{TeV}$ & PbPb & $\YnS{1}$ & $0.77$ & $0.39$ & $\WithUnit{37}{\%}$\\
		$\WithUnit{5.02}{TeV}$ & PbPb & $\YnS{2}$ & $0.22$ & $0.18$ & $\WithUnit{95}{\%}$\\
		\hline\hline
	\end{tabular}
\end{table}

Our results may be compared with those from related approaches to $\Y$-suppression such as~\cite{em12,striba12,peng11,son12}.
The model of Strickland and Bazow \cite{striba12} also includes dynamical propagation of the $\Y$~meson in the colored medium and a potential based on the heavy-quark internal energy.
The results are consistent with the STAR data for an initial central temperature of ${\WithUnit{428}{MeV} < T_0 < \WithUnit{442}{MeV}}$, whereas in the model of Liu et al. \cite{peng11} $T_0 = \WithUnit{340}{MeV}$ is used.
%Our result is in between these two approaches.
The strong binding model by Emerick, Zhao and Rapp \cite{em12} includes a contribution from cold nuclear matter effects and is also consistent with the STAR data.

The model by Song, Han and Ko \cite{son12} uses second-order gluon- and quark-dissociation of bottomia rather than first order as here and 
in other works such as \cite{peng11}. In-medium production and dissociation are calculated
from a rate equation. Wave functions and
decay widths are obtained from a screened Cornell potential that corresponds essentially to the real
part of the complex potential that we are using.
The fireball is modeled as a viscous, cylindrically symmetric fluid and
transversely averaged quantities are calculated. The inclusion of viscosity
allows for lower temperatures at the same QGP lifetime as compared to
perfect-fluid hydrodynamics in our modeling.
The two effects of bottomium regeneration and gluonic  (anti-) shadowing
are also included in the model, but are found to have only small
impact on the results. The model by Ko et al. does, however, not include an imaginary part in the potential
to account for the significant contribution of collisional damping to the total width, and the running of the
strong coupling $\alpha_s$ is not considered.

Some of the various model results are reviewed in comparison with recent data in \cite{an16}. Once the respective parameters are tuned, the results are often found to be compatible with the data in spite of vastly different model ingredients (such as different quark-antiquark potentials) and hence, it is difficult to extract model-independent conclusions. Regarding the relative importance of in-medium suppression and feed-down for ground and excited states as function of energy investigated in this work, our conclusions should, however, be quite stable, and it would be interesting to test this proposition in the other models.

%((add other model results at LHC energies))

\section{Conclusion}
In summary, we have investigated the suppression of the $\YnS{n}$ states in UU and PbPb~collisions at RHIC and LHC energies in a model that considers the in-medium processes gluodissociation, screening and damping.
The feed-down cascade from the excited bottomia states produces substantial additional ground-state suppression, since the excited states melt through screening, or depopulate through dissociation processes and hence, there is less feed-down to the $\YnS{1}$ ground state as compared to pp at the same energy.
In contrast, the suppression of the first excited state $\YnS{2}$ at both RHIC and LHC energies is largely due to the properties of the hot quark-gluon medium.

Our model results for the ground state are in agreement with the centrality-dependent STAR and CMS data~\cite{ver16,CMS-2012,jo15}.
The flat transverse-momentum dependence of the suppression factor is consistent with the preliminary CMS data when the relativistic Doppler effect due to the velocity of the moving bottomia relative to the expanding QGP together with the effect of feed-down reduction is properly considered. In minimum-bias $\WithUnit{193}{GeV}$~UU, only about $\WithUnit{20}{\%}$ of the total $\YnS{1}$ suppression $(1-\RAA)$ is due to the in-medium effects, whereas in $\WithUnit{2.76}{TeV}$~PbPb the in-medium contribution is already about $\WithUnit{30}{\%}$ and further increases in $\WithUnit{5.02}{TeV}$~PbPb.

The suppression of the first excited $\YnS{2}$ state which occurs mostly in the QGP-phase requires additional centrality-dependent dissociation mechanisms in the whole $p_T$-range, and in particular in very peripheral collisions.
Here the strong magnetic field caused by the moving spectators may in principle induce a centrality-dependent effect on both production and dissociation of the $\YnS{n}$ states.

Although the initial magnetic field is very short-lived, the field in the presence of a conducting quark-gluon medium decays on a time scale that is larger than the $\Y$ and $\chib$ formation time and comparable to the collision time scale~\cite{tu13,kha14}. However, its magnitude in the medium is considerably reduced and moreover, with increasing impact parameter in peripheral collisions, the largest effect is produced at intermediate centralities.
Hence the observed strong suppression of the $\YnS{2}$ state in very peripheral collisions can probably not be attributed to the magnetic field.

Instead, additional mechanisms which we have not yet accounted for, and which are acting differently on the ground and excited states are required. These may be provided by cold nuclear matter (CNM) effects, but nuclear shadowing is expected to be small \cite{son12}, and act on ground and excited states similarly.
Another possibility is hadronic dissociation -- mostly by the large number of pions in the final state even in more peripheral collisions -- which is likely to be relevant only for the excited bottomium states, but not for the strongly bound spin-triplet ground state.
In any case, the current modeling has to be refined if higher accuracy is desired.

For the centrality and transverse momentum dependence of $\YnS{1}$ we have made predictions at the LHC energy of $\WithUnit{5.02}{TeV}$~PbPb where results are currently being analyzed.
The ALICE Collaboration has released preliminary data for the $\YnS{1}$ centrality dependent yields in $\WithUnit{5.02}{TeV}$~PbPb \cite{had16} at rapidities $2.5 < |y| < 4.0$. 
Here the suppression is found to be slightly less than at $\WithUnit{2.76}{TeV}$, but almost compatible within the experimental error bars.
CMS has presented preliminary $\WithUnit{5.02}{TeV}$ data in the midrapidity region $|y| < 2.4$, but so far only for the double ratio $\RAAYnS{2}/\RAAYnS{1}$ \cite{kim16}. 
Once their suppression factors for the individual states are available, the consistency of the ALICE and CMS results, and the agreement with our prediction can be checked.
%For a direct comparison with our prediction at $\WithUnit{5.02}{TeV}$, however, the forthcoming CMS data in the mid-rapidity region $|y| < 2.4$ are needed.

\section*{Acknowledgements}
We are grateful to the STAR and CMS Collaborations for sharing their data, and to J.\,P.~Blaizot for discussions during his stay at the Heidelberg Institute for Theoretical Physics.
F.~Nendzig is now at Accso-Accelerated Solutions GmbH, Darmstadt.

This work has been partially supported by DFG through TRR\,33 at the Universities of Bonn, LMU Munich and Heidelberg.

%% The Appendices part is started with the command \appendix;
%% appendix sections are then done as normal sections
%% \appendix

%% \section{}
%% \label{}

%% If you have bibdatabase file and want bibtex to generate the
%% bibitems, please use
%%
%%  \bibliographystyle{elsarticle-num}
%%  \bibliography{<your bibdatabase>}
\bibliography{gw16}

%% else use the following coding to input the bibitems directly in the
%% TeX file.

%% \begin{thebibliography}{00}

%% \bibitem{label}
%% Text of bibliographic item

%% \bibitem{}

%% \end{thebibliography}
\end{document}